\documentclass[amsmath,amssymb,prc,final,showpacs,twocolumn]{revtex4-1}
 
\usepackage{bm,hyphenat,xspace}
\usepackage{graphicx,epsfig}

\newcommand {\mcu}{\mathcal{U}}

\newcommand {\mct}{\mathcal{T}}

\newcommand{\cm}{\mathrm{c\!\:\!.m\!\:\!.}}
\newcommand{\He}{{}^3\mathrm{He}}
\newcommand{\Hh}{{}^3\mathrm{H}}
\newcommand{\Hd}{{}^2\mathrm{H}}

\begin{document}

\title {Four-body calculation of elastic deuteron-deuteron scattering}
  
\author{A.~Deltuva} 
\affiliation
{Institute of Theoretical Physics and Astronomy, 
Vilnius University, A. Go\v{s}tauto 12, LT-01108 Vilnius, Lithuania
}

\author{A.~C.~Fonseca} 
\affiliation{Centro de F\'{\i}sica Nuclear da Universidade de Lisboa, 
P-1649-003 Lisboa, Portugal }

\received{29 June, 2015} 

\begin{abstract}
Fully converged calculations of deuteron-deuteron elastic scattering
observables are performed at energies above three- and four-body
breakup threshold. Differential cross sections and analyzing
powers are obtained using realistic nucleon-nucleon  force models
together with the Coulomb repulsion between protons.  For all observables we find a very
reasonable agreement with the available experimental data
limited to deuteron beam energies up to 25.3 MeV. 

\end{abstract}

 \maketitle

\section{Introduction \label{sec:intro}}
Four-nucleon reactions above breakup threshold pose a highly challenging
theoretical and computational problem. In this regime 
rigorous and realistic results have been obtained so far by  only two methodologies.
These are the complex-energy
method in the framework of momentum-space integral equations
\cite{deltuva:12c} and the complex scaling method in the 
framework of coordinate-space differential equations \cite{lazauskas:15a}.

In this work we continue our theoretical investigation of the
four-nucleon ($4N$) scattering above breakup threshold.
Previous studies 
\cite{deltuva:12c,deltuva:13c,deltuva:14a,deltuva:14b,deltuva:15c}
were mostly devoted to $4N$ reactions initiated
by nucleon-trinucleon ($N+3N$) collisions, i.e.,
neutron ($n$) or proton ($p$) beams impinging on
$\Hh$ or $\He$ targets. While $n+\Hh$ and $p+\He$ 
elastic scattering are dominated by states with total $4N$ 
isospin $\mct=1$, the coupled $n+\He$ and $p+\Hh$ reactions
involve both $\mct=0$ and $\mct=1$. Overall a good reproduction
of the experimental data was achieved when using realistic
nucleon-nucleon ($NN$) potentials. The most remarkable
discrepancies above breakup threshold are the nucleon analyzing power 
and polarization, 
especially for the $\Hh(p,n)\He$ charge exchange reaction, and the minimum
of the differential cross section in elastic $p+\He$ and
$n+\He$ scattering above 25 MeV nucleon energy.
One may raise the question whether these disagreements are 
dominated by $\mct=1$ states, or if $\mct=0$ components have similar
shortcomings as well. The study of deuteron-deuteron ($d+d$)
elastic scattering, the only $4N$ process dominated by
$\mct=0$ states, may shed some light on this issue.

Therefore in the present work we concentrate
 on  $d+d$ elastic scattering above breakup threshold.
In Ref.~\cite{deltuva:15a} we already calculated $d+d$ 
reactions around $E_d =10$ MeV deuteron beam energy and found quite
a good agreement with data, but also some inconsistencies
between different data sets, thereby calling for a more
extensive study over a wider energy range.
Here we present results at $E_d$ ranging from 
4.75 MeV, just slightly above the three-cluster threshold of
4.45 MeV, to 25.3 MeV. Restricting the model space to
$4N$ states with $\mct=0$ precludes to obtain simultaneously  reliable
amplitudes for $\Hd(d,p)\Hh$ and $\Hd(d,n)\He$ transfer reactions,
but has an important practical advantage, namely, allows for the
reduction of the number of basis states needed to get convergence for 
the $d+d$ elastic observables and thereby speeds up the calculations.

In Sec. II we present the theoretical framework and the reliability of using $\mct=0$
states alone to calculate  $d+d$ elastic scattering. Differential cross section and analyzing 
power results are shown in Sec. III and a summary is presented in Sec. IV.

\section{Theory \label{sec:eq}}

The four-particle collision process is described  by exact
Alt, Grassberger, and Sandhas (AGS) equations \cite{alt:jinr,grassberger:67} 
for the transition operators $\mcu_{\beta\alpha}$ whose
components are labeled according to the chains of partitions. 
Given that neutrons and protons in the isospin formalism are  
treated as identical particles,
there are only two chains of partitions  that
can be distinguished by the
two-cluster partitions, one ($\alpha =1$) being of the $3+1$ type, 
i.e., (12,3)4, and
another ($\alpha =2$) being of the $2+2$ type, i.e., (12)(34).
For the nucleon-trinucleon scattering in previous works 
we solved the symmetrized AGS equations for $\mcu_{\beta 1}$ but the 
reactions initiated by two deuterons require transition operators $\mcu_{\beta 2}$. 
In both cases the AGS equations share the same kernel but differ in the driving term.
Thus, in the present work we solve the integral equations
\begin{subequations} \label{eq:U}
\begin{align}  
\mcu_{12}  = {}&  (G_0  t  G_0)^{-1}  
 - P_{34}  U_1 G_0  t G_0  \mcu_{12} + U_2 G_0  t G_0  \mcu_{22} , 
\label{eq:U12} \\  
\mcu_{22}  = {}& (1 - P_{34}) U_1 G_0  t  G_0  \mcu_{12} . \label{eq:U22}
\end{align}
\end{subequations}
Here $t$ is the two-nucleon transition matrix, $U_1$ and $U_2$ are the
transition operators for  the 1+3 and 2+2 subsystems,
$P_{34}$ is the permutation operator of particles 3 and 4, and
$G_0 = (E+i\varepsilon-H_0)^{-1}$ is the free 
four-particle resolvent at the available energy $E$, whereas $H_0$ is the
free Hamiltonian.
Although the physical scattering process corresponds to $\varepsilon \to +0$, 
the complex energy method uses finite  $\varepsilon$ value when solving
the AGS equations numerically. The physical scattering amplitudes are then obtained 
by extrapolating finite  $\varepsilon$ results to the $\varepsilon \to +0$ limit.
The extrapolation procedure as well as the 
special method for  integrals with quasi-singularities encountered when solving
Eqs.~\eqref{eq:U} are described in detail in our previous works
\cite{deltuva:12c,carbonell:14a,deltuva:14b}.

\begin{figure}[!]
\begin{center}
\includegraphics[scale=0.6]{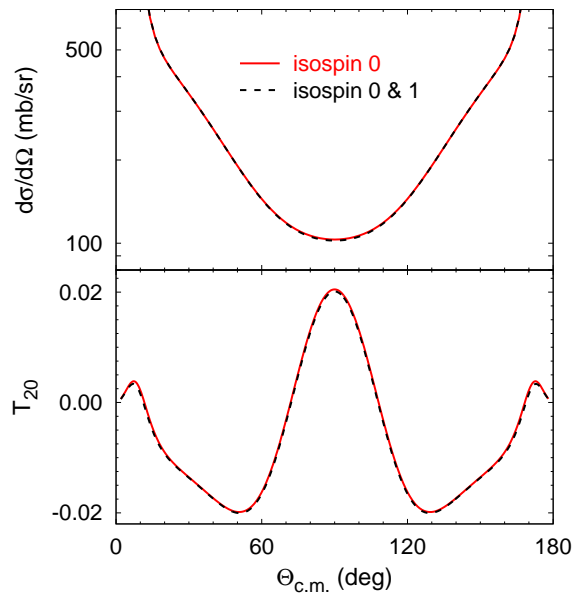}
\end{center}
\caption{\label{fig:iso} (Color online)
Deuteron-deuteron elastic scattering at 10 MeV beam energy.
Differential cross section and deuteron tensor analyzing 
power $T_{20}$ calculated neglecting (solid curves) and including (dashed curves)
total isospin $\mct=1$ states are compared.}
\end{figure}

The asymptotic  $d+d$ channel state is a pure total isospin $\mct=0$ state.
However, due to the charge dependence of the $NN$ interaction the 
scattering equations \eqref{eq:U} couple the states with different $\mct$.
Nevertheless, the effect of $\mct>0$ states on the $d+d$ elastic scattering 
is of second order in the charge-dependent terms of the $NN$ interaction
that are dominated by the $pp$ Coulomb force; a smaller contribution
from the hadronic part is present as well. Therefore at not too low energies
one may expect these effects to be small. This conjecture is well supported
by our test calculations for $d+d$ elastic scattering where we find a tiny effect of the 
$\mct=1$ states, as presented  in Fig.~\ref{fig:iso} for the differential cross section  $d\sigma/d\Omega$
and tensor analyzing power $T_{20}$, both as functions of the center-of-mass (c.m.) scattering angle
 $\Theta_\cm $.
 The results in the next section therefore are obtained by solving Eqs.~\eqref{eq:U} 
with $\mct=0$ states alone. This reduces the
 number of basis states  by more than a factor of 2
thereby speeding up the practical calculations significantly.
With  $\mct=0$ the isospin of the
$3N$ subsystem is limited to $T_y=\frac12$;
the needed isospin components of the 
two-nucleon transition matrix $t$ are given in Ref.~\cite{deltuva:14b}.)

The $pp$ Coulomb force is included using the method of screening and renormalization
\cite{alt:80a,deltuva:07b} where the screening radius $R = 10$ to 12 fm is found to be sufficient
to get convergence  for the Coulomb-distorted short-range part of the amplitude.
The obtained results are  well converged with respect to the partial-wave
expansion. When solving  Eqs.~\eqref{eq:U} we take into account isospin-singlet 
$2N$ partial waves
with  total angular momentum $j_x \leq 4$ and  isospin-triplet $2N$
partial waves  with orbital angular momentum $l_x \leq 7$, $3N$ partial
waves with spectator orbital angular momentum $l_y \leq 7$ and 
total angular momentum $J \leq \frac{13}{2}$, and $4N$ partial waves
with 1+3 and 2+2 orbital angular momentum $l_z \leq 7$. 
Initial and final deuteron-deuteron states with relative orbital angular momentum  
$L \leq 4$ are sufficient for the calculation of observables except
at the 25.3 MeV beam energy where we take into account also the states up to
$L \leq 6$ yielding a small but visible contribution.

\section{Results \label{sec:res}}

The scattering of two deuterons is
both challenging from the computational point of view and interesting
vis-a-vis nucleon-trinucleon scattering. Since deuterons are
loosely bound and spatially large objects, their collision gives rise
to much higher breakup cross sections than encountered in other   $4N$
reactions initiated by either neutrons or protons.

\begin{table}[htbp]
\begin{ruledtabular}
\begin{tabular}{l*{2}{c}}
 & $B(\Hh)$ & $B(\He)$    \\  \hline
N3LO        & 7.85 & 7.13 \\
CD Bonn     & 8.00 & 7.26  \\
CD Bonn + $\Delta$  & 8.28 & 7.53  \\
INOY04      & 8.49 & 7.73 \\
Experiment  & 8.48 & 7.72  
\end{tabular}
\end{ruledtabular}
\caption{$\Hh$ and $\He$ binding energies (in MeV)
for different $NN$ potentials.}
\label{tab:1}
\end{table}

\begin{figure*}[!]
\begin{center}
\includegraphics[scale=0.68]{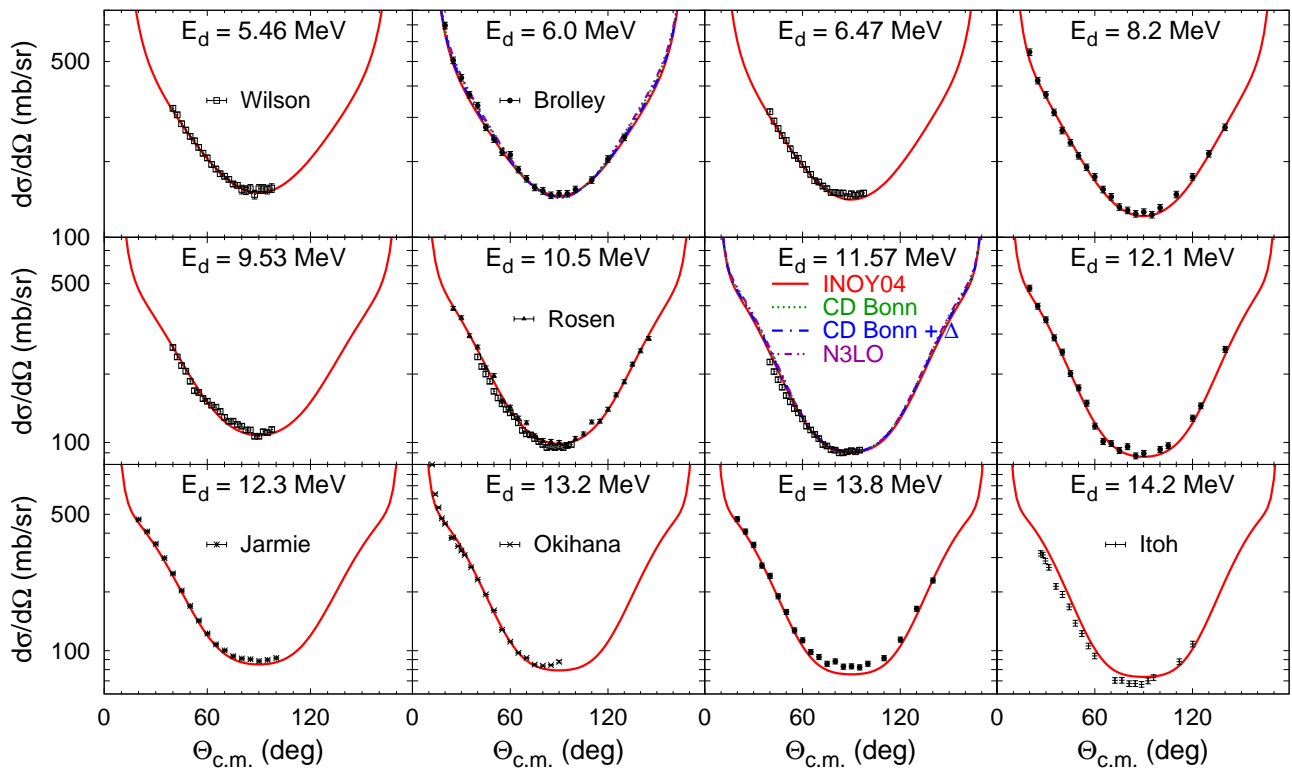}
\end{center}
\caption{\label{fig:s} (Color online)
Differential cross section of $d+d$ elastic scattering as a function
of c.m. scattering angle at 
deuteron beam energies ranging from 5.46 to 14.2 MeV.
Results are obtained using INOY04 potential (solid curves), and,
at 6.0 and 11.57 MeV, also CD Bonn + $\Delta$ (dashed-dotted 
curves), CD Bonn (dotted curves), and N3LO (double-dotted-dashed curves)
potentials. The experimental data  are from 
Refs.~\cite{wilson:69,brolley:60,rosen:52,PhysRevC.10.54,okihana:79,itoh:68}. }
\end{figure*}

\begin{figure}[!]
\begin{center}
\includegraphics[scale=0.68]{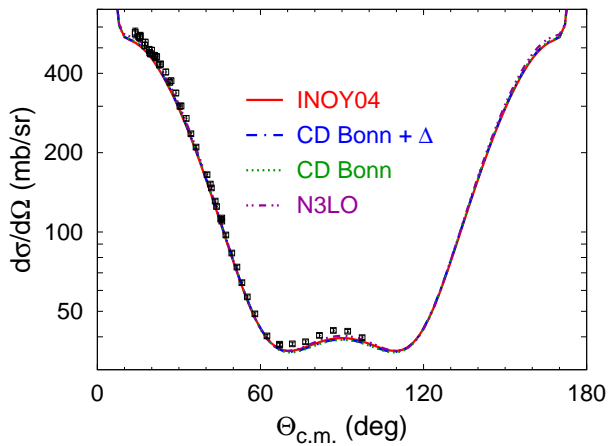}
\end{center}
\caption{\label{fig:s25} (Color online)
Differential cross section of $d+d$ elastic scattering
at $E_d = 25.3$ MeV. Curves are as in Fig.~\ref{fig:s}.
The experimental data are from Ref.~\cite{vanoers:63a}.}
\end{figure}

We calculate differential cross section and deuteron analyzing powers for
 $d+d$ elastic scattering at deuteron beam energies $E_d$ ranging from
4.75 to 25.3 MeV.  Given the identity of the deuterons, the observables are 
either symmetric or antisymmetric with respect to the center-of-mass scattering
angle $\Theta_\cm = 90^\circ$. 
At all considered energies the results are obtained using  
the realistic
inside-nonlocal outside-Yukawa (INOY04) potential  by Doleschall
\cite{doleschall:04a,lazauskas:04a}. It  nearly
reproduces the experimental values of $\He$  and $\Hh$
binding energy without an
additional $3N$ force. To investigate the dependence of the results on
the interaction model, at several energies,
i.e.,  $E_d = 6$, 10, 11.5, 11.57, and 25.3 MeV we show also the
predictions obtained with other high-precision $NN$ potentials.
These  are the  chiral effective field theory potential
at next-to-next-to-next-to-leading order (N3LO) \cite{entem:03a}, the
charge-dependent Bonn potential (CD Bonn)
\cite{machleidt:01a}, and its extension CD Bonn + $\Delta$
\cite{deltuva:03c} explicitly including an excitation of a nucleon to
a $\Delta$ isobar.  This mechanism generates  effective $3N$ and $4N$
forces that are  mutually consistent but quantitatively still
insufficient to reproduce $3N$ and $4N$ binding energies, although
they reduce the discrepancy \cite{deltuva:08a}.  The predictions
 of $\He$  and $\Hh$ binding energy for all employed force models
are collected in Table~\ref{tab:1}.

\begin{figure*}[!]
\begin{center}
\includegraphics[scale=0.62]{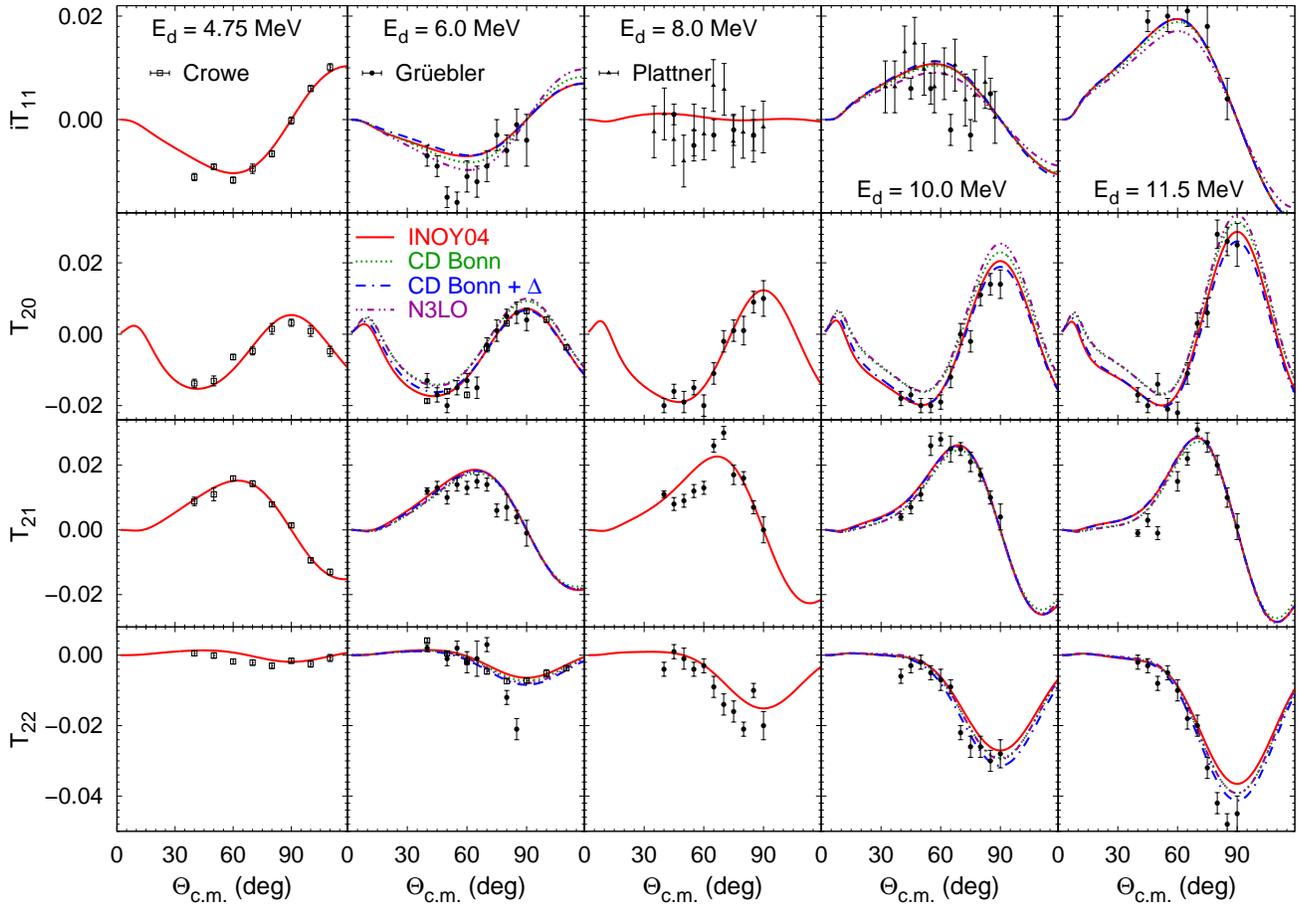}
\end{center}
\caption{\label{fig:t} (Color online)
Deuteron analyzing powers in $d+d$ elastic scattering at $E_d = 4.75$,
6.0, 8.0, 10.0, and 11.5 MeV. Curves are as in Fig.~\ref{fig:s}.
The experimental data are from 
Refs.~\cite{crowe:00a,gruebler:72b,plattner:69a}.}
\end{figure*}

\begin{figure}[!]
\begin{center}
\includegraphics[scale=0.6]{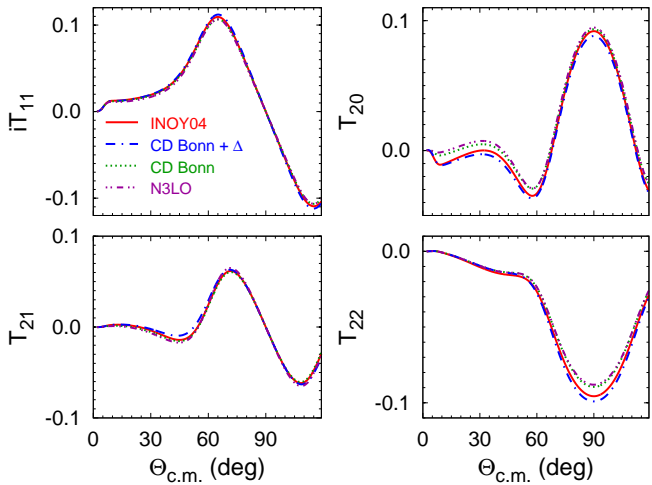}
\end{center}
\caption{\label{fig:t25} (Color online)
Deuteron analyzing powers in $d+d$ elastic scattering at $E_d = 25.3$
MeV. Curves are as in Fig.~\ref{fig:s}.}
\end{figure}

In Fig.~\ref{fig:s} we show results for the $d+d$ elastic differential cross
section in the $E_d$ range between 5.46 and 14.2 MeV. 
In this regime  $d\sigma/d\Omega$
shows quite a simple angular dependence, having  forward and backward 
peaks where $d\sigma/d\Omega$ becomes infinite due to the Coulomb force
and a single local minimum at  $\Theta_\cm = 90^\circ$. 
The calculations
describe the data \cite{wilson:69,brolley:60,rosen:52,PhysRevC.10.54} 
well up to $E_d = 12.1$ MeV and show
little sensitivity to the $NN$ force model as one may naively expect
given the large size of both projectile and target nuclei. 
The largest observed difference between the four employed force models
amounts about 6\% at $E_d = 6$ MeV around $\Theta_\cm = 40^\circ$ and  $140^\circ$.
For  $E_d$ above 12.3 MeV  small discrepancies between data and
calculations emerge, mostly around  $\Theta_\cm = 90^\circ$. 
However,  one probably should
question the quality of the data \cite{okihana:79} near $\Theta_\cm = 90^\circ$ where,
in contrast to all other sets,  a local maximum is present, 
as well as the 
quality of the data \cite{itoh:68} that lacks  symmetry with respect to
 $\Theta_\cm = 90^\circ$ and seems to be inconsistent with a smooth energy-dependence
seen in theoretical predictions and other data sets. Our calculations overpredict
the 14.2 MeV data \cite{itoh:68} by up to 9\% but  underpredict the 
13.8 MeV data  \cite{brolley:60}
around the minimum by the same amount.
Unfortunately, the available experimental data are very scarce above $E_d = 15$ MeV.
To the best of our knowledge, up to  $E_d = 35$ MeV which is the reach 
of the present calculations there is only one reliable data set at
$E_d = 25.3$ MeV \cite{vanoers:63a}. As shown in Fig.~\ref{fig:s25},
at this higher energy 
 the differential cross section finally develops a more complicated
angular dependence with a local maximum at  $\Theta_\cm = 90^\circ$
and two local minima around  $\Theta_\cm = 70^\circ$ and  $110^\circ$.
The calculations reproduce well the
shape of the data  but slightly underpredict its
magnitude by about 6\% at forward angles and around $\Theta_\cm = 90^\circ$.
Although the outlook is somehow contradictory, it looks like the calculations
slightly underpredict the experimental   $d\sigma/d\Omega$ data 
above $E_d=13$ MeV.
In contrast to lower energies where the minimum region is insensitive
to the $NN$ potential, at $E_d = 25.3$ MeV at $\Theta_\cm = 90^\circ$
the spread of predictions is about 3\% but shows no correlation with 
$3N$ binding energy.

In Fig.~\ref{fig:t} we  present the vector analyzing power $iT_{11}$ and
tensor analyzing powers $T_{20}$,  $T_{21}$, and   $T_{22}$ 
 for $E_d =  4.75$, 6.0, 8.0, 10.0, and 11.5 MeV.
Given the symmetry ($T_{20}$, $T_{22}$) or  antisymmetry ($iT_{11}$, $T_{21}$)
of these observables with respect to  $\Theta_\cm = 90^\circ$ and
the absence of the data for backward angles, we show only the regime 
up to  $\Theta_\cm = 120^\circ$. The overall agreement between 
theoretical predictions and experimental data is good.
Due to the large spatial size of the deuteron,
all these spin observables are very small in their magnitude, in a sharp contrast
with reactions involving initial and/or final nucleon-trinucleon states.
Unlike the differential cross section, the analyzing powers exhibit not
only a more complex behavior as  functions of energy  and scattering angle,
but also a greater sensitivity to the used $NN$ force model, especially
for  $T_{20}$ and around the extrema of  $iT_{11}$ and  $T_{22}$.
While tensor  analyzing powers monotonically increase in their magnitude with moderate
changes in the shape as the energy increases, 
the vector analyzing power $iT_{11}$ exhibits a change of sign after nearly vanishing
at $E_d= 8.0$ MeV. The error bars exceed the force model dependence for  $iT_{11}$
and  $T_{21}$.
Predictions based on N3LO and CD Bonn potentials slightly deviate from the data
for  $T_{20}$ above 10 MeV
whereas INOY04 and CD Bonn + $\Delta$ reproduce the data well. 
However, INOY04 slightly underestimates the magnitude of  $T_{22}$ 
whereas CD Bonn + $\Delta$  provides the best description of this observable.
In general, the dependence of the observables on the details of the interaction
is more complicated than just a simple scaling with the trinucleon binding
energy or deuteron $D$-state probability.
The $\Delta$-isobar excitation appears to be more important than
in the nucleon-trinucleon scattering.

  Although there is no data available for polarization observables
 at $E_d = 25.3$ MeV, we present them in Fig.~\ref{fig:t25} to demonstrate
the increase in their magnitude and the development of a more complicated
angular dependence. However, the relative sensitivity to the $NN$ force model
is not increased.

\section{Summary \label{sec:sum}}

In the present manuscript we solve the four-body AGS equations
for $d+d$ elastic scattering over a wide energy range above breakup
threshold using realistic force models between nucleons that are
based on either chiral effective field theory or meson exchange
theory. The Coulomb interaction between protons is included through
the method of screening and renormalization. The calculations are
fully converged in terms of partial wave expansion and discretization grids for 
the momentum variables.
In these calculation we
have only included total isospin $\mct=0$ states alone, given that
$\mct=1$ states have an extremely small contribution to $d+d$ elastic
scattering, as demonstrated in Sec. II.  Overall, no striking
disagreement with the data is observed. The calculated observables
follow the energy trend of the experimental data up to $E_d= 25.3$ MeV, which is
the maximum energy we have calculated at this time. We find that there
is a slight underprediction of the differential cross section minimum
beyond 13.2 MeV; however, this is not in a full analogy with the $p+\He$ and
$n+\He$ elastic scattering since the discrepancy in $d+d$ neither increases with 
the beam energy from 13.8 to 25.3 MeV  nor scales with $3N$ binding energy.
 We also observe no increase in sensitivity for spin
observables compared to what is found in $N+3N$ scattering.
In contrast to $N+3N$ and $p+d$ collisions,
the vector analyzing power in   $d+d$ elastic scattering 
is described quite well.



\end{document}